\documentclass{ws-ijmpa}
\usepackage{amsmath}
\usepackage{graphicx}
\usepackage[unicode,hidelinks]{hyperref}
\hypersetup{
  pdftitle={Fully three-dimensional L\'evy femtoscopic analysis of the pion pair source in UrQMD Au~+~Au collisions using a fast Fourier transform},
  pdfauthor={M\'aty\'as Moln\'ar}
}
\renewcommand{\wsdraftnote}{}

\graphicspath{{../figs/}{figs/}}
\newcommand{\snn}{\sqrt{s_{NN}}}
\newcommand{\kt}{k_{\mathrm T}}
\newcommand{\mt}{m_{\mathrm T}}
\newcommand{\rout}{R_{\mathrm{out}}}
\newcommand{\rside}{R_{\mathrm{side}}}
\newcommand{\rlong}{R_{\mathrm{long}}}
\newcommand{\rrms}{R_{\mathrm{RMS}}}
\newcommand{\rdiff}{R_{\mathrm{diff}}^2}
\newcommand{\vq}{\boldsymbol q}
\newcommand{\vrho}{\boldsymbol\rho}

\begin{document}
\markboth{M. Moln\'ar}{L\'evy source parameters from UrQMD}

\title{Fully three-dimensional L\'evy femtoscopic analysis\\
of the pion pair source in UrQMD Au~+~Au collisions\\
using a fast Fourier transform}
\author{M\'aty\'as Moln\'ar}
\address{Department of Atomic Physics, ELTE E\"otv\"os Lor\'and University,\\
P\'azm\'any P\'eter s\'et\'any 1/A, H-1117 Budapest, Hungary\\
molnarmatyas@student.elte.hu}
\maketitle

\begin{abstract}
Measurements of two-pion Bose--Einstein correlations motivate the use of
L\'evy-stable distributions to describe the non-Gaussian shape of pion
emission sources. In this work, we studied like-sign charged pion pairs in
0--10\% central Au+Au collisions simulated with UrQMD in hadronic cascade
mode at fifteen collision energies in the range $\snn=3$--$27$ GeV.
We compared two methods for extracting the L\'evy index $\alpha$ and the
three radii $\rout$, $\rside$, and $\rlong$. In the direct method, we
simultaneously fitted the out, side, and long projections of the
pair-separation distribution $D(\rho)$.
In the fast Fourier transform (FFT) method, we constructed the
quantum-statistical correlation function from the full distribution
and fitted it in three dimensions.
At fixed transverse mass, both methods yielded an overall decrease in
$\alpha$ and an increase in the radii with collision energy. Across the
matched bins, the FFT fits gave median reductions of 8\% in $\alpha$
and 15\% in $\rlong$ relative to the $D(\rho)$ fits. The FFT fits also
gave a larger $\rside$, while $\rout$ was similar in the two analyses.
In a separate $D(\rho)$ analysis at 3 GeV, the negative pion pairs tended
to give slightly larger $\alpha$ and $\rlong$ than the positive pairs,
with overlapping systematic uncertainties.
\end{abstract}


\section{Introduction}

Two-pion femtoscopy connects the structure of Bose--Einstein correlations
at small relative momentum to the space-time distribution of pion emission.
Experimental correlation measurements have found evidence for long tails or
pair sources better described by L\'evy-shaped distributions in several collision systems at the SPS, RHIC, and
the LHC.\cite{phenixlongtails,Adhikary2023SPS,Adare2018RHIC,Abdulameer2024RHIC,Tumasyan2024LHC,Kincses2024RHIC,CsanadKincses2024}
A symmetric L\'evy-stable distribution generalizes the Gaussian case
($\alpha=2$) to sources with power-law tails ($0<\alpha<2$).\cite{CsorgoHegyiZajc2004}
At a conjectured QCD critical point, the source index $\alpha$ has been
proposed to coincide with the spatial correlation exponent $\eta$.\cite{CsorgoCritical2006}
Hadronic rescattering and decays can also produce long tails, so a transport
baseline helps in interpreting the collision-energy dependence of
$\alpha$.\cite{Kincses2025,csanad2025investigating}

The UrQMD study in Ref.~\refcite{Molnar2026UrQMD} extracted a common $\alpha$
and three scale parameters by fitting the one-dimensional out, side, and long
projections of the pair-separation distribution $D(\rho)$. Experiments
measure the two-pion momentum correlation $C(\vq)$. In three-dimensional
analyses, they fit its dependence on the out, side, and long momentum
components simultaneously.\cite{STAR3D2005} A fast Fourier transform (FFT)
of the three-dimensional pair-separation distribution $D(\vrho)$ gives
access to the corresponding quantum-statistical correlation in UrQMD, allowing a comparison
with experiment in the same momentum variables. The three-dimensional fit
also samples directions between the axes, where sources with identical
one-dimensional projections can differ. We compared the $D(\rho)$ and FFT
methods to determine which trends survived this change of observable.
Because the main analysis combined the two pion charges, we also examined
the charge dependence of the $D(\rho)$ results separately at 3 GeV.

\section{Analysis methods}

\subsection{Simulated samples and pair kinematics}
\label{sec:kinematics}

The events were generated with \texttt{UrQMD 3.4} in hadronic cascade
mode.\cite{bass1998microscopic,bleicher1999relativistic,Molnar2026UrQMD}
The impact parameter was restricted to $0\le b\le 4.73$ fm, corresponding
approximately to the 0--10\% centrality class. The FFT analysis used
all $10^5$ events at each of the fifteen energies. The $D(\rho)$ analysis
used all $10^5$ events at $\snn=3.0$, 3.2, 3.5, 3.9, 4.2, 4.5, 5.2,
6.2, and 7.2 GeV, and the first $10^4$ at 7.7, 9.2, 11.5, 14.5,
19.6, and 27 GeV.
Same-event $\pi^+\pi^+$ and $\pi^-\pi^-$ pairs were pooled for the main
analysis. The track selection was $0.15\le p_{\mathrm T}<1.0$ GeV/$c$
and $|\eta|\le 1$. We defined $\kt$ as half the magnitude of the vector sum
of the two transverse momenta. The FFT analysis applied the explicit
cut $0.175\leq\kt\leq0.675$ GeV/$c$ and used ten bins of width
0.05 GeV/$c$. At eleven of the fifteen energies, we used the $D(\rho)$
fits of Ref.~\refcite{Molnar2026UrQMD}, whose first and last $\kt$ bins
were open-ended. At the remaining four energies (4.2, 5.2, 6.2, and
7.2 GeV), we repeated the $D(\rho)$ fits using the same procedure with
the explicit $\kt$ cut of the FFT analysis. The eight interior bins had
identical limits throughout.
The transverse mass is defined as $\mt=\sqrt{m_\pi^2+\kt^2}$
in units with $c=1$. We selected pairs with
$Q_{\mathrm{LCMS}}<\sqrt{A\mt}$, where $Q_{\mathrm{LCMS}}$
is the magnitude of the momentum difference evaluated in the longitudinally
co-moving system (LCMS). The cut selected pairs in the relative-momentum
range of the femtoscopic signal and allowed for its approximate
$\sqrt{\mt}$ width dependence.\cite{Molnar2026UrQMD}
We used $A=0.15$ GeV for the baseline selection.

The out axis follows the transverse momentum of the pair, the long axis follows
the beam, and the side axis completes the Bertsch--Pratt
basis.\cite{Pratt1986,Bertsch1988} Let $(\Delta t,\Delta\boldsymbol r)$
be the emission-coordinate difference after the boost into the LCMS,
where the longitudinal pair momentum vanishes. The effective separation
is then
\begin{equation}
  \vrho=(\Delta r_{\mathrm{out}}-\beta_{\mathrm T}\Delta t,\,
          \Delta r_{\mathrm{side}},\,\Delta r_{\mathrm{long}}),
  \qquad \beta_{\mathrm T}=\kt/K_{\mathrm{LCMS}}^0 ,
  \label{eq:rho}
\end{equation}
with $K_{\mathrm{LCMS}}^0$ the average pion energy in that frame.
The time term follows from the on-shell relation
$q^0=\beta_{\mathrm T}q_{\mathrm{out}}$ in the LCMS.
The pair density $D(\vrho)$ describes the distribution of this effective
separation in each $\kt$ bin.

\subsection{The $D(\rho)$ method}

For each pion pair, Eq.~(\ref{eq:rho}) gave the three components of its
effective separation $\vrho$ along the Bertsch--Pratt axes in the LCMS.
Within each $\kt$ bin, the pair contributed one entry to each of three
histograms, at $|\rho_{\mathrm{out}}|$, $|\rho_{\mathrm{side}}|$, and
$|\rho_{\mathrm{long}}|$, respectively. The out histogram, for example,
counted pairs according to their separation along the pair's transverse
momentum, including the emission-time correction. Summing over all side
and long separations in this way gave the out projection of $D(\vrho)$.
We used absolute values because interchanging the two pions reverses the
sign of each component.

We fitted the three histograms simultaneously with the corresponding
projections of a symmetric three-dimensional L\'evy source whose
principal axes were out, side, and long.\cite{Nolan2013LevyEq}
Each model projection has a one-dimensional L\'evy shape. The three
histograms shared the shape index $\alpha$ and normalization $\lambda^*$
in the fit. The radii $\rout$, $\rside$, and $\rlong$ set the length
scales along the respective axes. To account for the two signs
included in each histogram, we doubled the model density at positive
separations.

We examined the projections from 0.2 to 200.0 fm. The fit started at
3 fm, and its baseline upper limit was
$(80-5j)\sqrt{\snn/(11.5~\mathrm{GeV})}$ fm, where $j=0,\ldots,9$
labels the $\kt$ bins in increasing order. For each block of events,
we summed the pair counts and fitted all three projections with
the MIGRAD algorithm in Minuit2. We minimized the Poisson deviance
by comparing the observed and expected pair counts in each occupied
bin of the fit range. We then averaged the parameter values from the
accepted block fits, as described
in Ref.~\refcite{Molnar2026UrQMD}.

\subsection{The FFT method}

In the plane-wave approximation for identical pions, the squared
symmetrized wave function gives $1+\cos(\vq\cdot\vrho/\hbar c)$,
where $\vq=\boldsymbol p_1-\boldsymbol p_2$ is the full three-momentum
difference in the LCMS. We used the smoothness approximation, taking the
same normalized pair density for all $\vq$ within a given $\kt$ bin.
This gives the quantum-statistical
correlation\cite{Koonin1977,Pratt1984,Adare2018RHIC}
\begin{equation}
  C(\vq)=1+\int {\rm d}^3\rho\,
                  D(\vrho)\cos\!\left(\frac{\vq\cdot\vrho}{\hbar c}\right).
  \label{eq:fourier}
\end{equation}
Equation~(\ref{eq:fourier}) acts directly on the pair-separation density,
so $C(\vq)-1$ is its real Fourier transform. For independent emission
from a normalized single-particle source $S$, the pair density is an
autocorrelation, giving the familiar $|\widetilde S(\vq)|^2$ form.
The plane-wave kernel omits final-state interactions and thus gives the
Bose--Einstein contribution alone.

\begin{figure}[tbp]
  \centering
  \includegraphics[width=\linewidth]{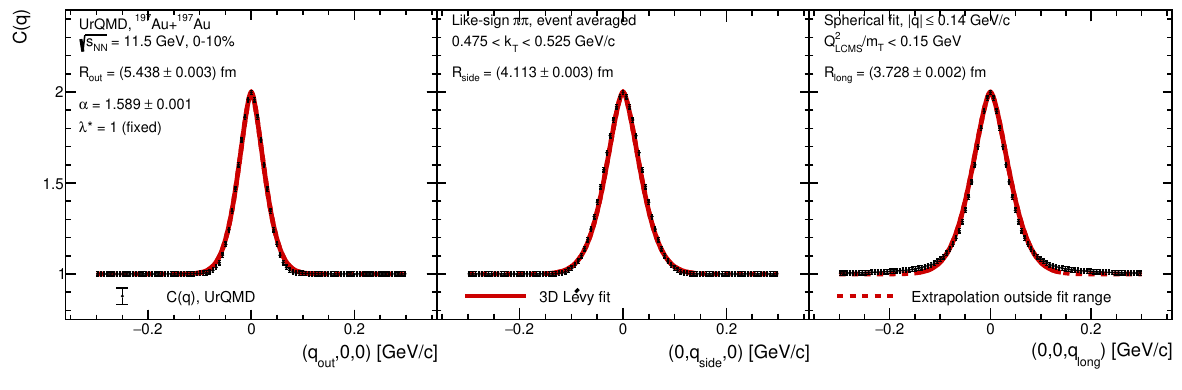}
  \caption{Central-axis slices of the FFT correlation and a single
  three-dimensional L\'evy fit at $\snn=11.5$ GeV and
  $0.475<\kt<0.525$ GeV/$c$, using the default pair selection. The fit included
  all bins within $|\vq|\le 0.14$ GeV/$c$. Along each axis, the solid
  line shows the fit within this range, and the dashed line shows its extrapolation.
  The point errors represent the common residual-based scale.}
  \label{fig:example}
\end{figure}

For each energy and $\kt$ bin, the signed pair coordinates were accumulated
over all events on a $251^3$ grid from $-100$ to $100$ fm on each axis.
The spatial bin width was about 0.80 fm, and the momentum spacing was
$2\pi\hbar c/(200~\mathrm{fm})\simeq 6.2$ MeV/$c$. We used \texttt{FFTW}
for the transform and accounted for the location of the first bin center
in the Fourier phase.\cite{FrigoJohnson2005}
The grid entries were normalized by the total pair count, including
the contents of the underflow and overflow bins. Thus, the numerical value $C(\boldsymbol 0)-1$
is the fraction of pairs retained in the grid. Across the 150 matched
bins this fraction ranged from 0.9727 to 1. In the example in
Fig.~\ref{fig:example}, it was 0.9968.

The function fitted to the full momentum-space histogram is
\begin{equation}
  C_{\mathrm{fit}}(\vq)=1+\lambda^*
  \exp\!\left[-\left(
       \sum_{i=\mathrm{out},\mathrm{side},\mathrm{long}}
          \frac{q_i^2 R_i^2}{(\hbar c)^2}
                   \right)^{\alpha/2}\right].
  \label{eq:levy}
\end{equation}
The model uses a diagonal L\'evy scale matrix and one index for all directions.
The fit included the origin, both signs of each momentum component, and
all bins in the sphere $|\vq|\le q_{\max}$. The baseline values of the fit limit
$q_{\max}$ were 0.10, 0.12, 0.14, and 0.16 GeV/$c$ for successive groups of three,
three, two, and two $\kt$ bins. This is a cut on the Fourier bins,
independent of the $Q_{\mathrm{LCMS}}$ selection on the pion pairs.

In the core--halo picture for chaotic emission, the observed correlation
strength is related to the squared fraction of pions emitted from the
resolvable core.\cite{Csorgo1996}
The cascade sample lacked pion feed-down from important long-lived parents
such as weakly decaying strange hadrons. The FFT grid covered only
$\pm100$ fm in each direction, and the $D(\rho)$ projections extended
to 200 fm. These finite spatial ranges limit sensitivity to a distant
halo. The core fraction cannot be inferred from these
fits. We therefore fixed $\lambda^*=1$, the core-only limit, in all FFT
fits. This sets the model intercept to $C_{\rm fit}(\boldsymbol 0)=2$,
while the intercept of the transformed histogram depends on the fraction
of pairs retained in the grid. The $D(\rho)$ fits used a floating
normalization.

We again used MIGRAD, this time to minimize the sum of squared
residuals with equal weight for every momentum bin in the fit sphere.
If MIGRAD missed its convergence criterion for the estimated distance
to the minimum, we retried once from its last parameter values, using
strategy 2 and a tolerance of $10^{-5}$ in place of $10^{-8}$.
We required convergence and a successful HESSE calculation with an
accurate covariance matrix for the four free parameters.
Following Ref.~\refcite{Kincses2026Geometry}, we rescaled this matrix using
the residual variance. Its denominator was the nominal number of degrees
of freedom, $N_{\rm bins}-4$. This count is only nominal because the
momentum bins are correlated. They are computed from the same source
counts, and each pair of mirror bins contributes twice to $N_{\rm bins}$
even though $C(\vq)=C(-\vq)$. The resulting curvature errors describe local fit
sensitivity. The rescaled $\chi^2$/NDF equals unity by construction.
The figures show the spread from the analysis variations.

\subsection{Systematic variations and matched bins}

For both methods, we varied $A$ to 0.05 and 0.25 GeV.
For $D(\rho)$, we also varied the upper spatial fit limit by $\pm25\%$
and the number of events in a block. The systematic uncertainties of
the original $D(\rho)$ analysis are given in Ref.~\refcite{Molnar2026UrQMD}.
For the FFT method, we varied the fit-sphere radius by $\pm20\%$ and
used 151 and 351 spatial bins per axis at the same spatial extent.
This varied the spatial discretization while leaving the momentum
spacing and the spatial cutoff unchanged.
Each variation changed one setting at a time. For a given source of
uncertainty, opposite shifts entered the two sides of the band
individually, while two shifts in the same direction were averaged on
that side. We combined the contributions in quadrature on each side.
Table~\ref{tab:fft-systematics} reports the mean relative contributions
over the ten $\kt$ bins at three representative energies. Across the scan, the pair selection gave
the largest contribution to both sides of the uncertainty for each
radius at every energy. The fit range dominated the upward uncertainty
on $\alpha$ throughout. The fit range also gave the largest downward
contribution to the uncertainty on $\alpha$ through 14.5 GeV. At 19.6
and 27 GeV, the pair-selection and fit-range contributions were comparable.

We matched results by collision energy and $\kt$ bin. All ten bins at
each of the fifteen energies had valid FFT fits for the baseline and
all six systematic variations.
For each method, we used the spread from its analysis variations to
estimate the systematic uncertainties shown by the bands.
The FFT curvature errors were generally smaller than the systematic
uncertainties. Their magnitude alone does not establish the statistical
precision of the comparison. Since the event samples overlap, a
statistical uncertainty on the difference would require the covariance
between the two estimates.

\section{Results}

\subsection{Transverse-mass dependence}

\begin{figure}[tbp]
  \centering
  \includegraphics[width=0.85\linewidth]{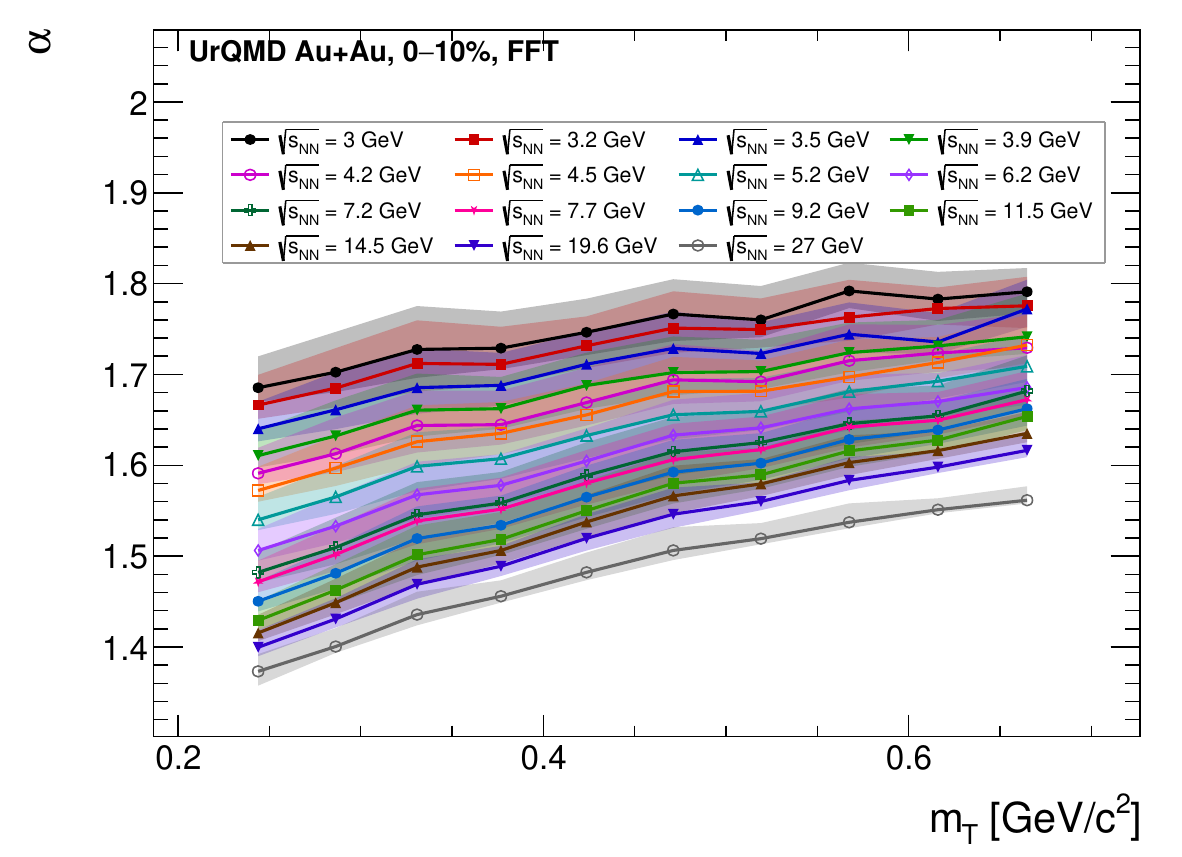}
  \caption{L\'evy index from the FFT fits as
  a function of $\mt$ at the fifteen energies. Lines connect the fitted
  values, and the shaded bands show systematic uncertainties.}
  \label{fig:mtalpha}
\end{figure}

The three slices in Fig.~\ref{fig:example} show how the directional
radii enter a single FFT fit. In this bin, the fitted peak was
narrowest along $q_{\mathrm{out}}$, corresponding to the largest
length scale, $\rout$, in Eq.~(\ref{eq:levy}).
Across the energy scan, the fitted $\alpha$ stayed below the Gaussian
value and generally increased with $\mt$ (Fig.~\ref{fig:mtalpha}).
Within the L\'evy parameterization, the smaller index at low $\mt$
corresponds to a more slowly falling source tail. The rise with $\mt$
was more pronounced at the higher energies. At the lowest energy, the
FFT fits still showed a rise, while $\alpha$ from the $D(\rho)$ fits
varied little across the measured range.\cite{Molnar2026UrQMD}

\begin{figure}[tbp]
  \centering
  \includegraphics[width=\linewidth]{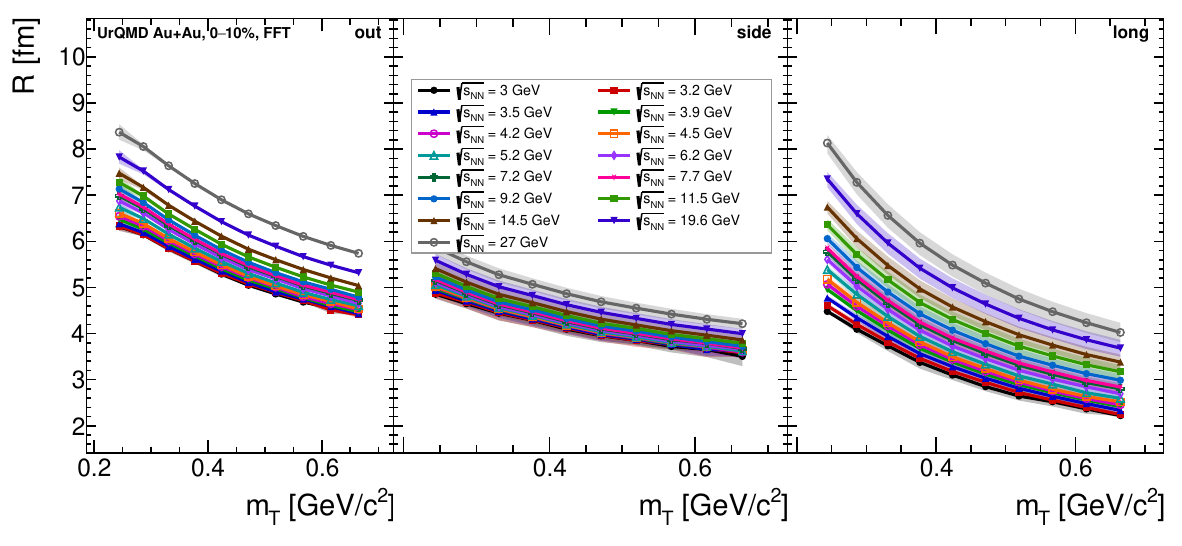}
  \caption{Out, side, and long L\'evy radii from the FFT fits as
  functions of $\mt$. Colors and systematic bands follow
  Fig.~\ref{fig:mtalpha}.}
  \label{fig:mtradii}
\end{figure}

All three FFT radii generally decreased with $\mt$
(Fig.~\ref{fig:mtradii}), as in the $D(\rho)$ analysis.\cite{Molnar2026UrQMD}
This behavior is consistent with position--momentum correlations in an
expanding source. Pions in a given momentum interval preferentially
sample regions with similar velocities, and the decreasing radii
indicate smaller regions of homogeneity for higher-$\kt$ pairs.

\subsection{Collision-energy dependence and comparison of methods}

\begin{figure}[tbp]
  \centering
  \includegraphics[width=0.85\linewidth]{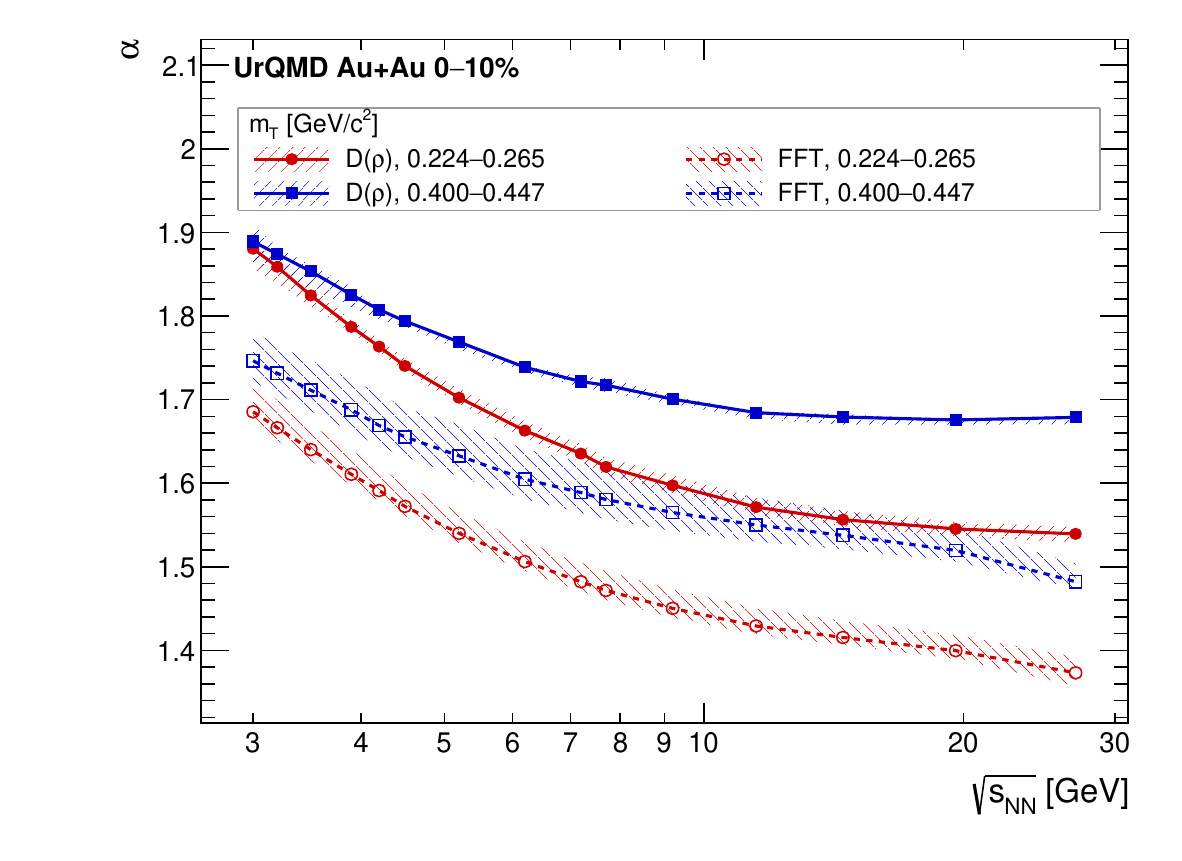}
  \caption{Collision-energy dependence of $\alpha$ from $D(\rho)$ and
  FFT fits in the intervals $0.224$--$0.265$ and $0.400$--$0.447$ GeV/$c^2$
  in $\mt$. The lower interval corresponds to the first $\kt$ bin,
  nominally $0.175$--$0.225$ GeV/$c$. At the eleven energies specified
  in Sec.~\ref{sec:kinematics}, this $D(\rho)$ bin also included pairs
  below 0.175 GeV/$c$. The higher interval had the same limits for both
  methods at every energy.
  Bands show systematic uncertainties. Lines guide the eye.}
  \label{fig:salpha}
\end{figure}

In both $\mt$ intervals shown in Fig.~\ref{fig:salpha}, $\alpha$
decreased overall with collision energy in both analyses. Preliminary
STAR measurements based on one-dimensional correlation fits also
showed a decrease with energy.\cite{STARCPOD2024,csanad2025investigating}

\begin{figure}[tbp]
  \centering
  \includegraphics[width=\linewidth]{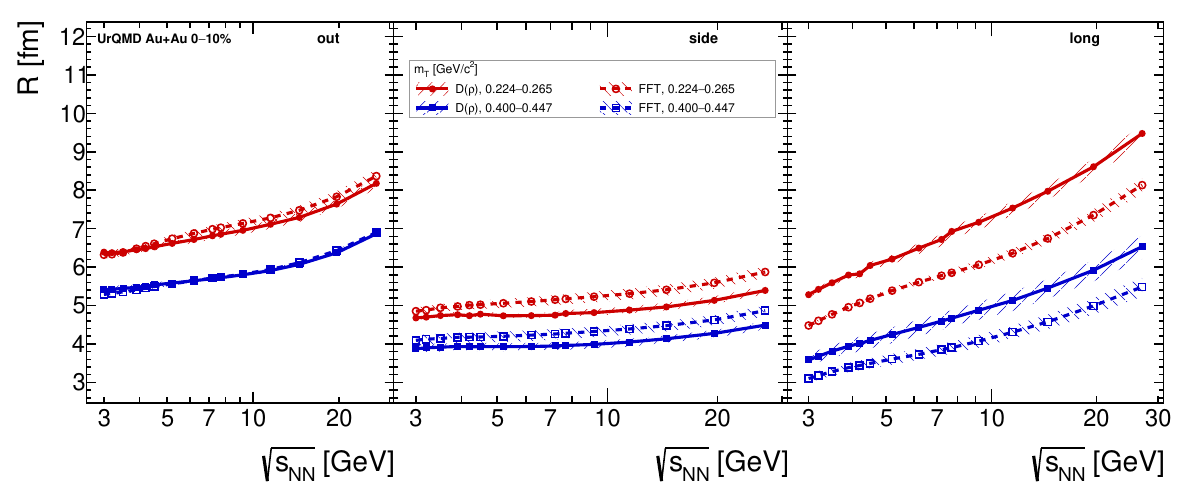}
  \caption{Collision-energy dependence of the out, side, and long
  L\'evy radii. The two $\mt$ intervals and the uncertainty convention
  are the same as in Fig.~\ref{fig:salpha}.}
  \label{fig:sradii}
\end{figure}

The three radii generally grew with energy in both $\mt$ intervals
(Fig.~\ref{fig:sradii}). In both analyses, the homogeneity region grew
most along the beam and least in the side direction, as seen in the
relative changes of $\rlong$ and $\rside$ across the scan.

In all 150 matched bins, including the edge bins, the FFT fits gave a
lower $\alpha$, a smaller $\rlong$, and a larger $\rside$ than the
$D(\rho)$ fits (Table~\ref{tab:comparison}). Restricting the comparison
to the 128 bins with identical $\kt$ acceptance gave median shifts of
$-7.9\%$ in $\alpha$ and $-14.8\%$ in $\rlong$, close to those for
the full sample. The two methods gave similar $\rout$, with overlapping
systematic bands in most bins. The opposite shifts in $\rside$ and
$\rlong$ partly canceled in $\rrms$, making it less sensitive
to the method than the longitudinal radius. This quantity summarizes
the fitted scales even when the second spatial moment of an ideal
L\'evy source diverges, as it does for $\alpha<2$.

\begin{table}[tbp]
  \renewcommand{\arraystretch}{1.2}
  \tbl{Relative shift $100(X_{\mathrm{FFT}}/X_{D(\rho)}-1)$ in the 150
  bins matched by collision energy and nominal $\kt$. Each bin contributes
  one value. The $\kt$ acceptance differed in 22 edge bins.
  Here $\rrms=\sqrt{(\rout^2+\rside^2+\rlong^2)/3}$ and
  $\rdiff=\rout^2-\rside^2$.}
  {\begin{tabular}{@{}lrrr@{}}
  \Hline
  Quantity & Minimum (\%) & Median (\%) & Maximum (\%)\\
  \hline
  $\alpha$          & $-13.09$ & $-7.97$  & $-3.69$\\
  $\rout$           & $-4.12$  & $+0.04$  & $+4.72$\\
  $\rside$          & $+3.76$  & $+8.01$  & $+12.27$\\
  $\rlong$          & $-17.34$ & $-14.63$ & $-6.16$\\
  $\rrms$           & $-4.15$  & $-2.50$  & $+2.62$\\
  $\rout/\rside$    & $-11.29$ & $-7.10$  & $-3.74$\\
  $\rdiff$          & $-34.98$ & $-14.71$ & $-4.94$\\
  \Hline
  \end{tabular}}
  \label{tab:comparison}
\end{table}

\begin{figure}[!htbp]
  \centering
  \includegraphics[width=0.78\linewidth]{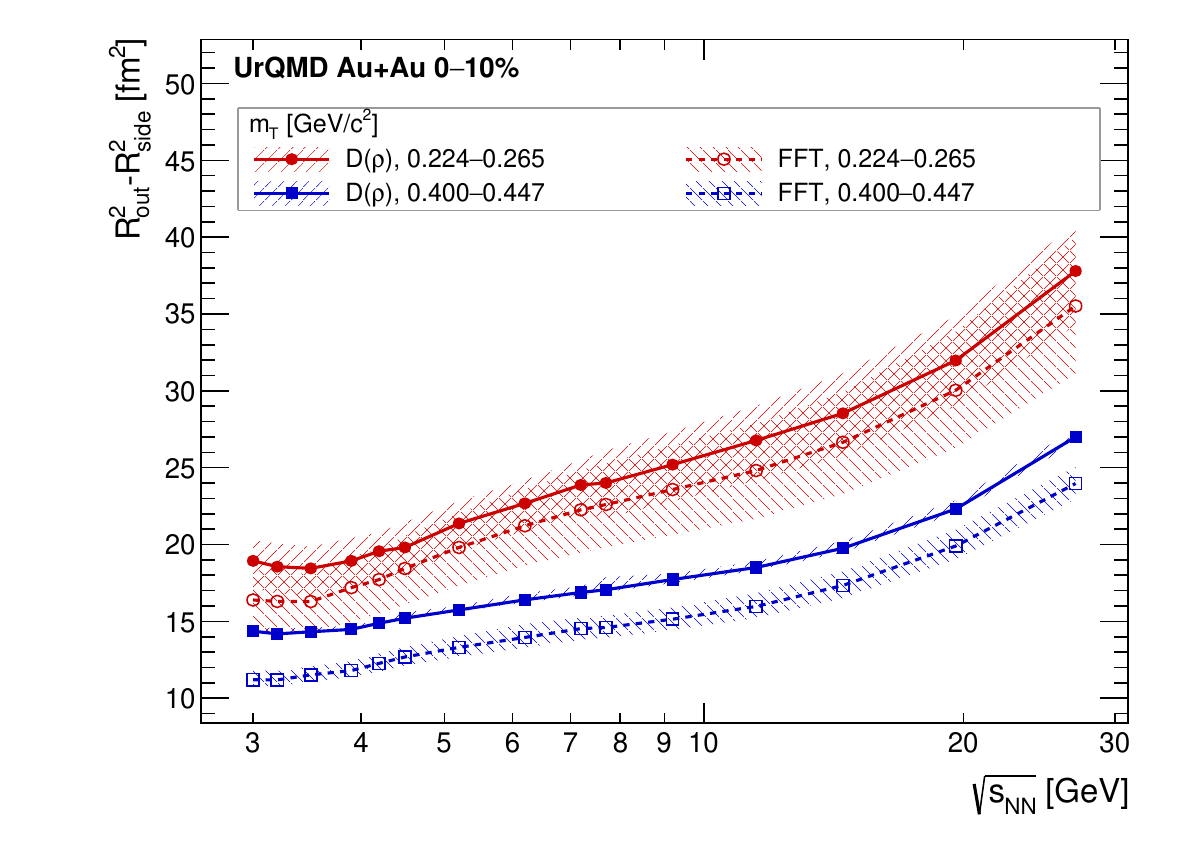}
  \caption{Difference of squared out and side L\'evy scales for both
  methods in the two $\mt$ intervals of Fig.~\ref{fig:salpha}.
  For each systematic variation, we formed $\rout^2-\rside^2$ from the
  two radii fitted together. The bands therefore retain their correlated
  response to each variation.}
  \label{fig:diff}
\end{figure}

The term $\beta_{\mathrm T}\Delta t$ in Eq.~(\ref{eq:rho}) makes the
out separation sensitive to the duration of emission. This motivates
examining $\rdiff$, although transverse geometry and correlations
between emission position and time also contribute.\cite{csanad2025investigating}
For $\alpha<2$, the L\'evy scales do not give finite source variances,
so the connection to emission duration is only qualitative.
The FFT fits brought the two transverse scales closer together.
Both $\rout/\rside$ and $\rdiff$ decreased relative to $D(\rho)$ in
every matched bin. The median reduction in $\rdiff$ was about 15\%,
while both methods retained its overall rise with collision energy
(Fig.~\ref{fig:diff}).

\begin{table}[!htbp]
  \setlength{\tabcolsep}{3pt}
  \renewcommand{\arraystretch}{1.04}
  \newcommand{\systpair}[2]{\ensuremath{+#1/{-}#2}}
  \tbl{Mean relative systematic uncertainties (\%) of the FFT fit parameters.
  Entries give the upward and downward changes averaged over $\kt$ bins.
  The total was formed in quadrature on each side within each bin before
  averaging. All ten bins contributed at each energy.}
  {\begin{tabular*}{\linewidth}{@{\extracolsep{\fill}}llcccc@{}}
  \Hline
  $\snn$ (GeV) & Source & $\alpha$ & $\rout$ & $\rside$ & $\rlong$\\
  \hline
  3.0 & $Q_{\rm LCMS}$ cut & \systpair{0.89}{0.85} & \systpair{0.81}{0.96} & \systpair{2.88}{3.29} & \systpair{2.65}{2.80}\\
   & $q_{\max}$ & \systpair{1.83}{0.96} & \systpair{0.16}{0.09} & \systpair{0.50}{1.16} & \systpair{1.10}{0.42}\\
   & Grid bins & \systpair{0.08}{0.02} & \systpair{0.11}{0.02} & \systpair{0.09}{0.06} & \systpair{0.37}{0.11}\\
   & Total & \systpair{2.10}{1.33} & \systpair{0.88}{0.97} & \systpair{2.93}{3.52} & \systpair{2.96}{2.89}\\
  \hline
  11.5 & $Q_{\rm LCMS}$ cut & \systpair{0.84}{0.68} & \systpair{0.45}{0.48} & \systpair{2.65}{2.79} & \systpair{3.58}{3.95}\\
   & $q_{\max}$ & \systpair{1.70}{0.81} & \systpair{0.26}{0.44} & \systpair{0.58}{1.45} & \systpair{1.20}{0.37}\\
   & Grid bins & \systpair{0.11}{0.03} & \systpair{0.07}{0.02} & \systpair{0.11}{0.03} & \systpair{0.22}{0.06}\\
   & Total & \systpair{1.91}{1.09} & \systpair{0.55}{0.69} & \systpair{2.72}{3.17} & \systpair{3.81}{3.97}\\
  \hline
  27.0 & $Q_{\rm LCMS}$ cut & \systpair{0.38}{0.37} & \systpair{0.45}{0.45} & \systpair{2.56}{2.73} & \systpair{4.08}{4.47}\\
   & $q_{\max}$ & \systpair{1.25}{0.37} & \systpair{0.12}{0.34} & \systpair{0.27}{1.01} & \systpair{0.57}{0.12}\\
   & Grid bins & \systpair{0.11}{0.03} & \systpair{0.04}{0.01} & \systpair{0.08}{0.02} & \systpair{0.15}{0.04}\\
   & Total & \systpair{1.32}{0.55} & \systpair{0.49}{0.64} & \systpair{2.58}{2.93} & \systpair{4.13}{4.47}\\
  \Hline
  \end{tabular*}}
  \label{tab:fft-systematics}
\end{table}

\clearpage
\subsection{Charge dependence at 3 GeV}

\begin{figure}[!htbp]
  \centering
  \includegraphics[width=0.60\linewidth]{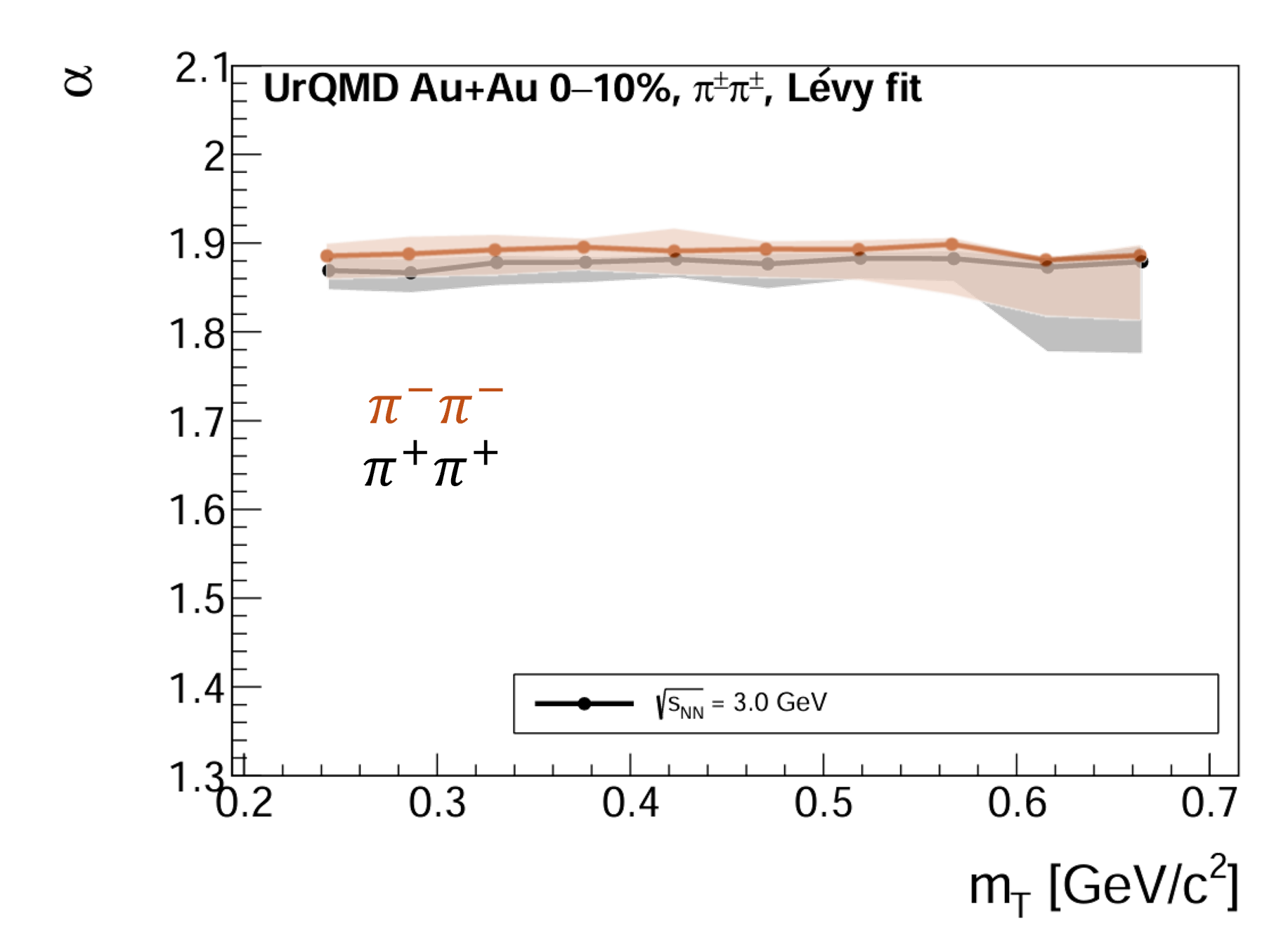}\\
  \includegraphics[width=\linewidth]{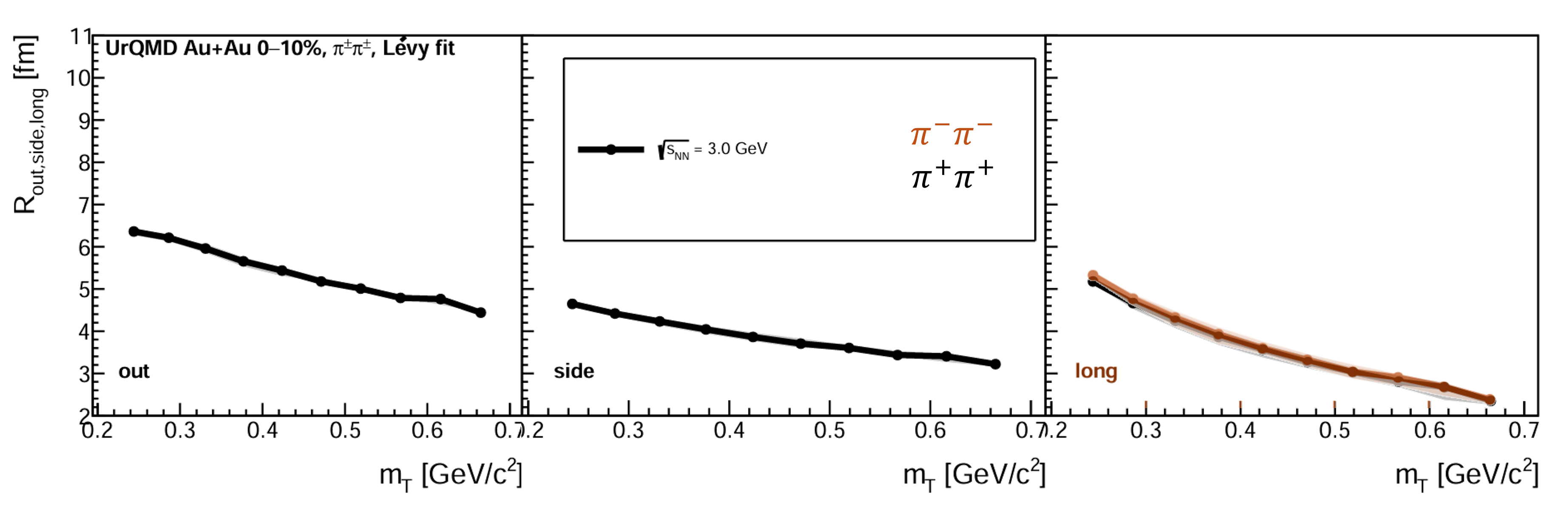}
  \caption{Separate $D(\rho)$ fits for $\pi^-\pi^-$ and $\pi^+\pi^+$ at
  $\snn=3$ GeV. The upper panel shows $\alpha$, and the lower panels
  show the three L\'evy scale parameters. Orange marks the negative-pair
  result in the $\alpha$ and $\rlong$ panels. The shaded bands indicate
  systematic uncertainties.}
  \label{fig:charge}
\end{figure}

We examined the $D(\rho)$ results for $\pi^+\pi^+$ and $\pi^-\pi^-$
separately at $\snn=3$ GeV. Figure~\ref{fig:charge} shows a small upward
shift for the negative pairs in $\alpha$ and, most visibly among the
radii, in $\rlong$. At $\mt\simeq0.42$ GeV/$c^2$, for example, the
negative-pair index exceeded the positive-pair index by about 0.01.
The systematic bands for the two charges overlapped. STAR reported
larger $\pi^-\pi^-$ side and long Gaussian radii before correction for
Coulomb interaction with the net charge of the emitting system. After
this correction, the radii agreed within uncertainties.\cite{Qi2026STAR}

The $D(\rho)$ calculation used the generator's freeze-out coordinates.
It therefore omitted both the pair Coulomb wave function and the
post-freeze-out momentum shift from the residual charged system.
Charge-dependent hadronic production and rescattering offer another
possible contribution. In particular, UrQMD calculations at 7.7 GeV
have identified initial-state isospin as a possible source of a small
pion charge splitting even in the absence of a residual Coulomb
field.\cite{KhyzhniakLisa2026} Establishing a charge splitting in this
sample would require the uncertainty on the charge difference, including
the correlated response to the common analysis variations.

\section{Discussion and conclusions}

At fixed $\mt$, both methods found that $\alpha$ generally decreased
and the radii grew with collision energy, most strongly along the beam.
The radii also generally decreased with increasing $\mt$. At the same
energy and $\mt$, the FFT fits gave a lower $\alpha$, a larger $\rside$,
and a smaller $\rlong$, while $\rout$ changed little. Two features of
the comparison could account for these shifts. The three projections
discard information about the joint distribution of separations, and
the two fits weight different parts of the source and correlation peak.

Consider the out momentum axis, where $q_{\rm side}=q_{\rm long}=0$. Each pair
then contributes $\cos(q_{\rm out}\rho_{\rm out}/\hbar c)$ to the
integral in Eq.~(\ref{eq:fourier}). The sign of $\rho_{\rm out}$ drops
out because the cosine is even. The $|\rho_{\rm out}|$ histogram thus
contains all the separations needed to calculate $C(q_{\rm out},0,0)-1$,
provided we use the same pairs and normalization in both calculations.
Away from the axes, the cosine depends on two or three separation
components of the same pair. The three separate histograms lose the
information about which out, side, and long separations occurred together.

To test how much this distinction could matter, we constructed a toy source
with independent out, side, and long separations. The three projections
were L\'evy distributions with a common $\alpha=1.6$ and respective
radii of 6, 5, and 5 fm.
Its correlation, $C(\vq)-1=\exp[-\sum_i |q_i R_i/(\hbar c)|^\alpha]$,
agrees with Eq.~(\ref{eq:levy}) on each axis for $\lambda^*=1$, but its
peak falls more steeply between the axes. A fit to the three axes
recovered the input parameters. We then fitted Eq.~(\ref{eq:levy}) to
the entire cube $|q_i|<0.3$ GeV/$c$, with equal weights, a grid
spacing of 5 MeV/$c$, and $\lambda^*$ fixed at unity. To describe the
narrower peak between the axes, the fit increased all three radii by
about 9\%, while $\alpha$ changed only from 1.600 to 1.597. The extra
information between the axes in this particular source thus produced
the opposite shift in $\rlong$ and far too small a change in $\alpha$
to account for the UrQMD results.

The fit ranges and weights offer another possible explanation. Even a
source with ellipsoidal contours can have a central peak and a tail
that a single L\'evy index describes only approximately. The $D(\rho)$
fit counts pairs over a restricted interval of separations, starting
at 3 fm. In the FFT fit, pairs at widely different separations
contribute to the same momentum bin. Long separations influence the
narrow structure near $\vq=0$, where there are relatively few bins
in a uniformly weighted three-dimensional fit. The two fits can
therefore favor different indices and radii even for this geometry.
Event averaging and normalization also differed. The $D(\rho)$ analysis
averaged fits to event blocks with a floating normalization. The
FFT analysis fitted one pooled histogram with $\lambda^*$ fixed at
unity. From 7.7 GeV upwards, that histogram also contained ten times
as many events as the $D(\rho)$ fit sample. The present comparison
measures the combined effect of these
choices. Fitting just the three axes of the same UrQMD correlation,
with the momentum range, bin weights, and amplitude prescription held
fixed, would show how much the fitted parameters move when the other
momentum bins are included.

Both methods reproduced the same overall energy dependence, but the FFT
fits yielded median shifts of about $-8\%$ in $\alpha$ and $-15\%$ in
$\rlong$ relative to the $D(\rho)$ fits. For a quantitative comparison
with a three-dimensional measurement, the model correlation should be
fitted in the same momentum variables and under comparable fit conditions.

In the separate $D(\rho)$ fits at 3 GeV, negative pion pairs tended to
give slightly larger $\alpha$ and $\rlong$ than positive pion pairs,
with overlapping systematic bands.

\section*{Acknowledgments}

The author acknowledges the support of the National Research, Development and Innovation Office (NKFIH, Hungary) research grant 2026-4.1.1-MISSZIÓK-2026-00005 and NKKP ADVANCED 152097.

\section*{Data availability}

The systematic uncertainty estimates for the FFT fits and their
breakdown by source at all fifteen energies are available on
Zenodo (\href{https://doi.org/10.5281/zenodo.24567481}{doi:10.5281/zenodo.24567481}).

\bibliographystyle{ws-ijmpa}
\bibliography{wpcf2026_references}

@article{phenixlongtails,
  author = {Adler, S. S. and others},
  collaboration = {PHENIX},
  title = {Evidence for a Long-Range Component in the Pion Emission Source in {Au+Au} Collisions at $\sqrt{s_{NN}}=200$ {GeV}},
  journal = {Phys. Rev. Lett.},
  volume = {98},
  pages = {132301},
  year = {2007},
  doi = {10.1103/PhysRevLett.98.132301}
}

@article{Adhikary2023SPS,
  author = {Adhikary, H. and others},
  collaboration = {NA61/SHINE},
  title = {Two-pion femtoscopic correlations in {Be+Be} collisions at $\sqrt{s_{NN}}=16.84$ {GeV} measured by the {NA61/SHINE} at {CERN}},
  journal = {Eur. Phys. J. C},
  volume = {83},
  pages = {919},
  year = {2023},
  doi = {10.1140/epjc/s10052-023-11997-8}
}

@article{Adare2018RHIC,
  author = {Adare, A. and others},
  collaboration = {PHENIX},
  title = {L\'evy-stable two-pion {Bose--Einstein} correlations in $\sqrt{s_{NN}}=200$ {GeV} {Au+Au} collisions},
  journal = {Phys. Rev. C},
  volume = {97},
  pages = {064911},
  year = {2018},
  doi = {10.1103/PhysRevC.97.064911}
}

@article{Abdulameer2024RHIC,
  author = {Abdulameer, N. J. and others},
  collaboration = {PHENIX},
  title = {Centrality dependence of L\'evy-stable two-pion {Bose--Einstein} correlations in $\sqrt{s_{NN}}=200$ {GeV} {Au+Au} collisions},
  journal = {Phys. Rev. C},
  volume = {110},
  pages = {064909},
  year = {2024},
  doi = {10.1103/PhysRevC.110.064909}
}

@article{Tumasyan2024LHC,
  author = {Tumasyan, A. and others},
  collaboration = {CMS},
  title = {Two-particle {Bose--Einstein} correlations and their L\'evy parameters in {PbPb} collisions at $\sqrt{s_{NN}}=5.02$ {TeV}},
  journal = {Phys. Rev. C},
  volume = {109},
  pages = {024914},
  year = {2024},
  doi = {10.1103/PhysRevC.109.024914}
}

@article{Kincses2024RHIC,
  author = {Kincses, D.},
  title = {Pion interferometry with L\'evy-stable sources in $\sqrt{s_{NN}}=200$ {GeV} {Au+Au} collisions at {STAR}},
  journal = {Universe},
  volume = {10},
  pages = {102},
  year = {2024},
  doi = {10.3390/universe10030102}
}

@article{CsanadKincses2024,
  author = {Csan\'ad, M. and Kincses, D.},
  title = {Femtoscopy with L\'evy sources from {SPS} through {RHIC} to {LHC}},
  journal = {Universe},
  volume = {10},
  pages = {54},
  year = {2024},
  doi = {10.3390/universe10020054}
}

@article{Kincses2025,
  author = {Kincses, D. and Nagy, M. and Csan\'ad, M.},
  title = {L\'evy walk of pions in heavy-ion collisions},
  journal = {Commun. Phys.},
  volume = {8},
  pages = {55},
  year = {2025},
  doi = {10.1038/s42005-025-01973-x}
}

@article{csanad2025investigating,
  author = {Csan\'ad, M. and Kincses, D.},
  title = {Investigating the excitation function of {HBT} radii for L\'evy-stable sources},
  journal = {J. Phys. G},
  volume = {52},
  pages = {025102},
  year = {2025},
  doi = {10.1088/1361-6471/ada04f}
}

@article{Molnar2026UrQMD,
  author = {Moln\'ar, M. and Kincses, D. and Huang, Y. and Csan\'ad, M.},
  title = {Pion emission source shape in {UrQMD} {Au+Au} collisions at {STAR} energies},
  journal = {Eur. Phys. J. A},
  volume = {62},
  pages = {208},
  year = {2026},
  doi = {10.1140/epja/s10050-026-01970-w}
}

@article{bass1998microscopic,
  author = {Bass, S. A. and others},
  title = {Microscopic models for ultrarelativistic heavy ion collisions},
  journal = {Prog. Part. Nucl. Phys.},
  volume = {41},
  pages = {255--369},
  year = {1998},
  doi = {10.1016/S0146-6410(98)00058-1}
}

@article{bleicher1999relativistic,
  author = {Bleicher, M. and others},
  title = {Relativistic hadron--hadron collisions in the ultra-relativistic quantum molecular dynamics model},
  journal = {J. Phys. G},
  volume = {25},
  pages = {1859--1896},
  year = {1999},
  doi = {10.1088/0954-3899/25/9/308}
}

@article{Pratt1986,
  author = {Pratt, S.},
  title = {Pion interferometry of quark-gluon plasma},
  journal = {Phys. Rev. D},
  volume = {33},
  pages = {1314},
  year = {1986},
  doi = {10.1103/PhysRevD.33.1314}
}

@article{CsorgoHegyiZajc2004,
  author = {Cs\"org\H{o}, T. and Hegyi, S. and Zajc, W. A.},
  title = {{Bose--Einstein} correlations for {L\'evy} stable source distributions},
  journal = {Eur. Phys. J. C},
  volume = {36},
  pages = {67--78},
  year = {2004},
  doi = {10.1140/epjc/s2004-01870-9}
}

@article{CsorgoCritical2006,
  author = {Cs\"org\H{o}, T. and Hegyi, S. and Nov\'ak, T. and Zajc, W. A.},
  title = {{Bose--Einstein} or {HBT} correlation signature of a second order {QCD} phase transition},
  journal = {AIP Conf. Proc.},
  volume = {828},
  pages = {525--532},
  year = {2006},
  doi = {10.1063/1.2197465}
}

@article{STAR3D2005,
  author = {Adams, J. and others},
  collaboration = {STAR},
  title = {Pion interferometry in {Au+Au} collisions at $\sqrt{s_{NN}}=200$ {GeV}},
  journal = {Phys. Rev. C},
  volume = {71},
  pages = {044906},
  year = {2005},
  doi = {10.1103/PhysRevC.71.044906}
}

@article{Bertsch1988,
  author = {Bertsch, G. and Gong, M. and Tohyama, M.},
  title = {Pion interferometry in ultrarelativistic heavy-ion collisions},
  journal = {Phys. Rev. C},
  volume = {37},
  pages = {1896--1900},
  year = {1988},
  doi = {10.1103/PhysRevC.37.1896}
}

@article{Nolan2013LevyEq,
  author = {Nolan, J. P.},
  title = {Multivariate elliptically contoured stable distributions: theory and estimation},
  journal = {Comput. Stat.},
  volume = {28},
  pages = {2067--2089},
  year = {2013},
  doi = {10.1007/s00180-013-0396-7}
}

@article{Koonin1977,
  author = {Koonin, S. E.},
  title = {Proton pictures of high-energy nuclear collisions},
  journal = {Phys. Lett. B},
  volume = {70},
  pages = {43--47},
  year = {1977},
  doi = {10.1016/0370-2693(77)90340-9}
}

@article{Pratt1984,
  author = {Pratt, S.},
  title = {Pion interferometry for exploding sources},
  journal = {Phys. Rev. Lett.},
  volume = {53},
  pages = {1219--1221},
  year = {1984},
  doi = {10.1103/PhysRevLett.53.1219}
}

@article{FrigoJohnson2005,
  author = {Frigo, M. and Johnson, S. G.},
  title = {The design and implementation of {FFTW3}},
  journal = {Proc. IEEE},
  volume = {93},
  pages = {216--231},
  year = {2005},
  doi = {10.1109/JPROC.2004.840301}
}

@article{Csorgo1996,
  author = {Cs\"org\H{o}, T. and L\"orstad, B. and Zim\'anyi, J.},
  title = {{Bose--Einstein} correlations for systems with large halo},
  journal = {Z. Phys. C},
  volume = {71},
  pages = {491--497},
  year = {1996},
  doi = {10.1007/BF02907008}
}

@unpublished{Kincses2026Geometry,
  author = {Kincses, D. and Karmakar, B. and Karpenko, I. and Tom\'a\v{s}ik, B.},
  title = {Nuclear Geometry-Driven Femtoscopic Correlations in Light-Ion Collisions},
  note = {arXiv:2610.29760 [nucl-th], Supplemental Material},
  year = {2026},
  doi = {10.48550/arXiv.2610.29760}
}

@misc{STARCPOD2024,
  author = {Kincses, D.},
  title = {Beam-energy dependent pion interferometry with {L\'evy}-stable sources at {STAR},},
  howpublished = {presentation for the {STAR} Collaboration, {CPOD} 2024},
  url = {https://drupal.star.bnl.gov/STAR/files/kincsesd_CPOD2024_talk_v6.pdf}
}

@article{Qi2026STAR,
  author = {Qi, Y.},
  collaboration = {STAR},
  title = {Third-body-corrected two-pion femtoscopic correlations in {Au+Au} collisions at high baryon density},
  journal = {J. Subatomic Part. Cosmol.},
  volume = {6},
  pages = {100531},
  year = {2026},
  doi = {10.1016/j.jspc.2026.100531}
}

@unpublished{KhyzhniakLisa2026,
  author = {Khyzhniak, Y. and Lisa, M. A.},
  title = {When positive and negative pairs differ in femtoscopy: residual {Coulomb} and isospin effects},
  note = {arXiv:2606.06472 [nucl-th]},
  year = {2026},
  doi = {10.48550/arXiv.2606.06472}
}
\end{document}